# Chemical Guidance in the Search for Past and Extant Life on Mars

Steven A. Benner[a]
Elisa Biondi[a]
Hyo-Joong Kim[a]
Jan Špaček[b]

[a]Foundation for Applied Molecular Evolution
13709 Progress Boulevard, Alachua FL (USA)

[b]Department of Biophysical Chemistry and Molecular Oncology
Institute of Biophysics of the Czech Academy of Sciences
Královopolská 135, 612 65 Brno, (Czech Republic)

NASA should design missions to Mars for the purpose of generating "Aha!" discoveries to jolt scientists contemplating the molecular origins of life. These missions should be designed with an understanding of the privileged chemistry that likely created RNA prebiotically on Earth, and universal chemical principles that constrain the structure of Darwinian molecules generally.

**Background**

Research to understand the origin of life is central to the NASA mission, and has been recognized as such for a half century. However, as many have noted, because of the antiquity of that event, laboratory research is poorly constrained by hard data. As a result, the field is characterized by the pursuit of idées fixes, half-hypotheses, and scripted reaction sequences, often for decades. Before his untimely death, Robert Shapiro likened such sequences to "a golfer who had successfully played a golf ball around an 18 hole course, [assuming] that the ball could then play the same course by itself, through a combination of the wind and other natural forces."[1] Further, scientists working on "origins" do not have easy access to "Aha!" discoveries that might productively jolt them from established lines of thinking. Such jolts are, of course, extremely useful in all other sciences.

In particular, laboratory research for how life originated in the Solar System lacks strong constraints concerning the relevant environments at the relevant times on the relevant planet. If that planet is Earth, the oldest physical records come as zircons ~4.35 billion years old. These may entrap elements (e.g. $Ce^{3+}/Ce^{4+}$) that indicate the redox states of fluids that they crystallized from,[2] trace elements (e.g. boron) that are proposed to constrain maturation of carbohydrates possibly relevant to RNA-first models for the origin of life,[3] and carbon, whose $^{12}C/^{13}C$ ratio might provide hints of a biosphere.[4] Most of the rest of the physical record on earth before 4+ Ga has been destroyed by plate tectonics.

This physical record is "not nothing". Further, it has been possible to integrate the scarce terran physical record with models for Solar System evolution, histories of planetary bombardment,[5] and models for planetary evolution late in accretion. To this is added analysis of late arriving materials in meteorites and limited samples from the Moon and other Solar System bodies.

From this, we can arrive at privileged chemistry that almost certainly occurred in the redox-neutral atmosphere that most likely dominated in the Hadean. This included formaldehyde (perhaps $3 \times 10^8$ molecules formed per $cm^2$ every second in the Hadean atmosphere)[6], and trace amounts of glycolaldehyde (perhaps one part per million of formaldehyde)[7]. These materials could not help have matured by well-known chemistry,[8] especially if the reactive aldehydes were delivered to the Hadean surface as bisulfite addition products generated from sulfur dioxide[9] that was certainly emerging from the mantle if the $Ce^{3+}/Ce^{4+}$ redox states are correctly inferred.

Unfortunately, the standard redox-neutral atmosphere is very unproductive with respect to the hydrogen cyanide, cyanamide, cyanogen, and cyanoacetylene that are the staples of much laboratory prebiotic research.[10] However, models for planetary accretion in late impact events do offer the possibility that the Hadean atmosphere was transiently reducing.[11] This reducing atmosphere would have transiently generated hydrogen cyanide, cyanamide, cyanogen, and cyanoacetylene (a good model is the modern Titan) that would allow generation of nucleobases found in modern RNA. The length of time for a productive atmosphere to complete prebiotic RNA synthesis depends on the size of the impactor. A Moon-sized body ($10^{23}$ kg, Moneta) would have been big enough to deliver the late veneer heavy metals (like platinum and gold) that arrived on Earth after Earth's core closed,[12] and would have created a cyanide-generating atmosphere that would have lasted for tens of millions of years. However, Moneta was very sterilizing; it would have created a lava ocean, and would likely have re-set most of the geological clocks on Earth.

A smaller $10^{21}$ kg Ceres-sized impactor would also have been sterilizing; it would not however have reset all of the clocks that we find on Earth. However,, but it would have delivered a productively reducing atmosphere that lasted perhaps one million years. A still smaller $10^{20}$ kg Vesta-sized impactor would not have sterilized the Earth or reset the clocks, but still would have generated a productively reducing atmosphere, albeit for just a few tens of thousands of years.

As a result, a nearly continuous model of chemistry leads directly from atmosphere-generated molecules (formaldehyde and trace glycolaldehyde in in all atmospheres, and reduced nitrogen - containing molecules in a post-impact atmosphere) all the way to oligomeric RNA. This relies on the presence however of phosphite[13] and/or cyclic trimeta-phosphate mineral species that are generated either from the impact or a reduction neutral mantle.[14] Simple paths are now known to generate nucleoside triphosphates and diphosphates,[15] and silica phases are now known to create oligomeric RNA from those triphosphates and diphosphates.[16] Further, assumptions about the impact history of the planet Earth provide estimates of the date when this most likely occurred.[17] Of course, RNA has a repeating backbone charge and size interchangeable building blocks. It is well known to support Darwinian evolution.

**Mars**

Remarkably, these models for the prebiotic chemistry that leads directly to the formation of RNA on Earth are entirely consistent with environments that we especially to have been present on early Mars. In particular, borate has been found on the evaporite surface of Mars today. Oceans were present on the surface of Mars at the time when there was life on Earth. Indeed, since dry land is necessary for any model for the origin of RNA that relies on atmospheric input, and since the amount of dry land on early Earth is not constrained, but could be quite low,[18] Mars might even be a better place for models that move continuously from atmospheric input to oligomeric RNA by way of borate, phosphate, and other minerals. That is, the same processes may have generated life on Mars and, in some parameters are more likely to have generated life on Mars.

This generates two mission proposals:
(a) Search for extant life on Mars
(b) Search for evidence on Mars of past conditions where these continuous models operate.

Missions to search for extant life

The key challenge in any search for a biosignature is uncertainty about what molecules to search for. It is difficult to reproduce on Mars the full power of an analytical chemistry that we can establish on earth. Therefore, we cannot simply directing NASA mission planners to "look for everything".

Fortunately, laboratory studies over the past two decades have sought to synthesize alternative systems that might support Darwinian evolution. This work constrains the *kinds* of molecules that could support alien genetics. These constraints in turn constrain how we must build mission instruments to search for life on Mars. Two sets of constraints based on two principles have been particularly robust.

First, the principal suggested by Erwin Schrödinger 1943 has been fully corroborated by laboratory experiments. To support faithful replication, the interchangeable building blocks must all fit into an "aperiodic crystal" structure, ensuring that mutations that change the encoding information do not change the size and shape of the replicating molecule. This allows the physics of phase transitions to ensure accurate genetic information transfer.

The polyelectrolyte theory of the gene provides a second principle, analogous to the first. Here, genetic the universal genetic molecule is not constrained to have the same size and shape upon evolution, but required to have the same properties upon evolution. This requires that molecules directly encoding evolving information do not change dramatically their physical properties as their information content changes. In water, this is conveniently achieved by having those building blocks assembled on a backbone that has a repeating charge. In terran informational molecules (DNA and RNA) that repeating charge is negative. However, theory and experiments allow backbones with a repeating positive charge as a possible alien genetic information molecule.

These theories constrain the structures of the molecules that prebiotic chemistry must generate to support the first Darwinian evolution, a process that is presumed to be necessary for life universally. Specifically, prebiotic processes must generate polyelectrolytes built from a privileged set of building blocks that all have the same size and shape. Further, these theories constrain the structures of molecules that must be sought any mission to Mars to seek extant life. These are, again all electrolytes built from a privilege that the building blocks of all of the same size and shape.

Fortunately, polyelectrolytes are easily concentrated from dilute solution. Further, various classes of analysis can be envisioned that would determine whether or not those polyelectrolytes are built from a privileged, constraints of the building blocks all having similar size and shape. Such molecules are unlikely to arise without the influence of Darwinian evolution, since the number of conceivable building blocks is astronomical and covers all shapes and sizes. Here, it is important to note that homochirality is itself a kind of uniform size and shape, something required for the Schrödinger aperiodic crystal structure.

**The first proposal for this white paper therefore is to have NASA seek creative proposals to concentrate from water on Mars polyelectrolytes and to analyze them to see whether they conform to Schrödinger's aperiodic crystal structure.**

Missions to search for conditions where past life could have emerged

However, unlike Earth, Mars has not suffered plate tectonics; none of its record has been lost by subduction. Therefore, the Martian surface including recent volcanism is likely to accurately reflect in general terms the minerals that were available to possibly originate life in the Hadean (Noachian).

Here, mission design must reflect the fact that Mars has nevertheless not been static over the last 3 to 4 billion years. Photochemistry followed by the gravitational escape of dihydrogen means that the accessible surface on Mars today is likely far more oxidized than in the Hadean (Noachian). This has implications for the recent generation of organic species, as well as for the survival of ancient organic species. Further, the ancient Martian hydrosphere has fractionated and transformed many of its mineral species.

Even so, recent rovers have detected borate and magnesium containing minerals that are integral to the continuous model for the formation of prebiotic RNA. Further, the Martian atmosphere continues to generate formaldehyde, although very small amounts. It is conceivable that the same processes in the continuous model for the prebiotic origin of RNA are still operating on Mars today, should water become from time to time available. Mars mission design should consider this fact.

However, even in the absence of extant life, are understanding of privilege chemistry and carbon dioxide-rich atmospheres intermittently reduced by impacts, above a redox neutral mantle offers exact prescriptions for Mars mission design relative to the past capability of Mars to generate RNA. **From this we can generate several proposals for the design of missions to Mars and, in some cases, the Moon:**

1. A better understanding of the subsurface accessible by volcanism. Mars has had recent volcanism, making accessible to Martian missions igneous rocks that can be examined to test the redox potential of the mantle.[19] This should be a target for geological examination of the surface.
2. The late veneer. Various siderophiles (e.g. Au, Pt) were delivered to the Earth by late impacts after the core had closed; core closure removed most of these with iron into the center of the Earth. A comparison of the veneer on the Earth and the veneer on the Moon is central to our effort to evaluate the number and size of late impactors to Earth. However, those numbers have large experimental errors, even on Earth, but also on the Moon and Mars. Making better assessments of this material should be a target for the geological examinations of the surface of all planetary bodies, but especially the Moon and Mars.
3. The continuous model for the prebiotic formation of RNA occurs mostly in constrained aquifers that are evaporating. Here, the composition of the evaporites is critical. Making better assessments of the composition of those evaporites on Mars should be a prime focus of future missions so that planets. In particular, assessment of the abundance of phosphorus, boron, carbon, and sulfur is critical.
4. Entrapped organic materials.. As noted above, the Martian surface has become increasingly oxidizing due to photochemical action the gravitational escape of dihydrogen. This means that organic materials are unlikely to survive, especially in the presence of hydroxyl radicals, unless they are especially stable to further oxidation.[20] However, past missions have shown that such organics might be protected by oxidation by the chloride ion, due to its ability to be oxidized to perchlorate, which itself is a kinetically poor oxidant at low-temperature. Future missions to Mars should search for these entrapped organics in such stabilized environments.